\documentclass[11pt]{article}
\pdfoutput=1

\usepackage{jheppub}
\usepackage{rotating}

\usepackage{wrapfig}
\usepackage[left=5cm,right=0cm,top=5cm,bottom=1cm,footskip=1cm]{geometry}

\usepackage{amsfonts}
\usepackage{amsmath}
\usepackage{slashed}
\usepackage{amssymb}
\usepackage{latexsym}
\usepackage{multirow}
\usepackage{hyperref}
\usepackage{enumitem}
\usepackage{mathtools}
\hypersetup{colorlinks=true,citecolor=red,linkcolor=magenta,urlcolor=blue}
\usepackage{relsize}
\usepackage{booktabs}
\usepackage{subfigure}
\usepackage{cleveref}
\usepackage{fancyhdr}
\usepackage{float}
\usepackage{graphicx}
\usepackage{microtype}
\usepackage{tabularx}
\usepackage{xcolor}
\usepackage{siunitx,adjustbox,bm}
\usepackage[bottom]{footmisc}

\usepackage{physics}

\renewcommand{\arraystretch}{1.3}

\newcolumntype{C}{>{$}c<{$}}
\newcolumntype{L}{>{$}l<{$}}
\newcolumntype{R}{>{$}r<{$}}

\title{Effective 2HDM Yukawa Interactions and a Strong First-Order Electroweak Phase Transition}
\author[a]{Anisha,}
\author[b]{Duarte Azevedo,}
\author[b]{Lisa Biermann,}  
\author[a]{Christoph Englert,} 
\author[b]{and Margarete M\"uhlleitner}
\affiliation[a]{School of Physics \& Astronomy, University of Glasgow, Glasgow G12 8QQ, United Kingdom}
\affiliation[b]{Institute for Theoretical Physics, Karlsruhe Institute of Technology, 76128 Karlsruhe, Germany}
\emailAdd{anisha@glasgow.ac.uk}
\emailAdd{duarte.azevedo@kit.edu}
\emailAdd{lisa.biermann@kit.edu}
\emailAdd{christoph.englert@glasgow.ac.uk}
\emailAdd{margarete.muehlleitner@kit.edu}

\abstract{The top quark as the heaviest particle in the Standard Model
  (SM) defines an important mass scale for Higgs physics and the
  electroweak scale itself. It is therefore a well-motivated degree of
  freedom which could reveal the presence of new interactions beyond
  the SM. Correlating modifications of the top-Higgs interactions in
  the 2-Higgs-Doublet Model (2HDM), we analyse effective field theory
  deformations of these interactions from the point of view of a
  strong first-order electroweak phase transition (SFOEWPT). We show that such modifications are compatible with current Higgs data and that an SFOEWPT can be tantamount to a current overestimate of exotic Higgs searches' sensitivity at the LHC in $t\bar t$ and four top quark final states. We argue that these searches remain robust from the point of accidental signal-background interference so that the current experimental strategy might well lead to 2HDM-like discoveries in the near future.}

\begin{document}
\maketitle
\allowdisplaybreaks
\section{Introduction}
\label{sec:intro}
The lack of direct evidence for new interactions beyond the Standard Model (BSM) at the Large Hadron
Collider (LHC) and other experiments is puzzling given the theoretically and experimentally established 
need to go beyond the Standard Model (SM). As experiments are moving increasingly towards
model-independent methods to report measurements and BSM sensitivity, a range of established BSM
phenomena continue to signpost particular sectors of the SM for further phenomenological scrutiny. One such sector
is related to the interactions of the top quark. The top quark, as the heaviest believed-fundamental particle enters
a range of phenomenologically accessible final states at the LHC. It decays before hadronisation thus enabling 
the direct analysis of its properties, whilst abundantly produced in hadronic collisions. Furthermore, it creates
a large radiative pull of electroweak interactions, which is highlighted by the sensitivity of the electroweak fit
to the top mass~\cite{Baak:2012kk}, the metastability of the electroweak vacuum at high scales~\cite{Degrassi:2012ry,Bednyakov:2015sca}, as well as, its role
as a threshold in Higgs physics~(e.g.~\cite{Baur:2002rb,Englert:2013vua}). It might be fair to say that the ``right'' theory of BSM interactions seems
further away from discovery than ever, but the top quark and its relation to the weak scale might well hold the key
to unlocking the secrets of electroweak symmetry breaking.\footnote{This is echoed by the central part that the 
top plays in concrete models of BSM physics, ranging from supersymmetry to strongly interacting models.}

The critical role of the top quark is apparent from its strong coupling to the Higgs field with a Yukawa interaction of order unity in the SM.
The qualitative pattern predicted by the SM has been spectacularly validated by the discovery of the Higgs boson in $H\to \gamma\gamma$
decays with direct evidence from top-associated Higgs production providing mounting evidence of the SM-like character of top-Higgs interactions,
alluding to fundamental mass generation for the top quark through the electroweak vacuum. This relation also puts the top
quark centre-stage for the emergence of the non-trivial vacuum itself in the early history of our Universe, potentially playing
{\emph{the}} critical role in facilitating a strong first-order electroweak phase transition (SFOEWPT) in the context of electroweak baryogenesis to address the Sakharov
criteria~\cite{Sakharov:1967dj} for matter anti-matter asymmetry. Additional sources of CP violation (under the assumption that baryogenesis  
proceeds canonically) could be traceable into phases of the Yukawa interactions (for recent phenomenological analyses see Refs.~\cite{Englert:2019xhk,Bortolato:2020zcg,Barman:2021yfh}), 
and their appearance is indicative of a richer scalar sector such as the 2-Higgs-Doublet Model (2HDM)~\cite{Lee:1973iz,Branco:2011iw,Fontes:2017zfn,Basler:2018dac}. To this end, in this work, we focus on the
possibility of obtaining an SFOEWPT in the 2HDM with a specific focus on the role of the top
quark. The characteristics of additional SFOEWPT-relevant contributions in the scalar sector have been discussed in Refs.~\cite{Anisha:2022hgv}, highlighting a qualitative
agreement with similar discussions in the context of the SM: Additional Higgs interactions that lead to an SFOEWPT show up predominantly as modifications
of Higgs pair interactions via modifications of the Higgs self-coupling. Following the canonical arguments of thermodynamics, such modifications should
predominantly be visible in the phenomenology of the light degrees of freedom, in agreement with the findings of Ref.~\cite{Anisha:2022hgv}. In flavon extensions of the SM, 
it has been observed that large Yukawa coupling modifications can lead to the desired SFOEWPT~\cite{Baldes:2016rqn}. When these effects are captured by the top
quark modifications, this can lead to large departures from the expected phenomenology of the BSM states.

In this paper, we consider a motivated effective field theory (EFT) extension of the top quark sector in the 2HDM. On the one hand, this addresses the emerging tension
of observing an SFOEWPT in the 2HDM of type II given the current Higgs coupling measurements~\cite{Basler:2016obg,Atkinson:2021eox,Atkinson:2022pcn,Atkinson:2022qnl}; on the other hand the means of EFT enable us
to remain agnostic to the particular extension of the 2HDM.\footnote{Employing matching computations~\cite{DasBakshi:2018vni,Carmona:2021xtq,Fuentes-Martin:2022jrf,Cohen:2020fcu,Dawson:2022cmu,Dawson:2023ebe}, results can then be connected to concrete UV extensions of the 2HDM. We will not discuss this further in this work. It is furthermore worth noting that considering non-SM degrees as dynamical rather than turning directly to SMEFT is particularly motivated given the limitations that SMEFT faces when considering electroweak phase transitions~\cite{Postma:2020toi}.}

This paper is organised as follows: Section~\ref{sec:2hdmeft} gives an
overview of the 2HDM type II and its EFT extension relevant to this
work. Section~\ref{sec:EPT} details the relevance of these
modifications for an SFOEWPT which is backed up by a comprehensive
scan over the 2HDM's type II EFT extension. As these results are
relevant for the phenomenology programme at the LHC, we perform a
detailed analysis of the EFT modifications for Higgs physics as a
function of an SFOEWPT in Sec.~\ref{sec:results}. We conclude in
Sec.~\ref{sec:conc}.

\section{The 2HDM and its Dimension-6 Yukawa Extension}
\label{sec:2hdmeft}
We start our discussion with the canonical 2HDM dimension-4 Yukawa terms which are given by 
by~\cite{Gunion:1989we,Gunion:2002zf}
\begin{equation}
\label{eq:2HDM-Yukawa}
	\mathcal{L}^{\text{dim-4}}_{\text{Yuk}} =- Y^e_{1}\bar{L}\Phi_1 e - Y^e_{2}\bar{L}\Phi_2 e - Y^d_{1}\bar{Q}\Phi_1 d- Y^d_{2}\bar{Q}\Phi_2 d   - Y^u_{1}\bar{Q}\tilde{\Phi}_1 u - Y^u_{2}\bar{Q}\tilde{\Phi}_2 u + \text{h.c.}\,,
\end{equation}
where $\Phi_{1,2}$ are $SU(2)_L$ doublets with hypercharge $Y=1$. The two doublets are expanded as
\begin{equation}
	\Phi_{1}= \begin{pmatrix}
	\phi^{+}_{1} \\
	\frac{1}{\sqrt{2}}(v_{1}+\zeta_{1}+i \psi_{1})
		\end{pmatrix} \,, \hspace{1.5cm} 
		\Phi_{2}=\begin{pmatrix}
		\phi^{+}_{2} \\ 
		\frac{1}{\sqrt{2}}(v_{2}+\zeta_{2}+i \psi_{2})
	\end{pmatrix} \,.
	\label{eq:higgsdoublets}
\end{equation}
Here, $v_{1}$ and $v_{2}$ are the vacuum expectation value (vev) of
$\Phi_{1}$ and $\Phi_{2}$, respectively, with $v_{1}^2 +v_{2}^2 = v^2$
and $v\simeq246 \;\text{GeV}$.  The $\phi_{i}^+$ label the charged fields and $\zeta_{i}$ is the neutral CP-even and $\psi_{i}$ is the neutral CP-odd field for $i=1,2$.
Motivated by the possibility of connecting the 2HDM of type II to high-scale supersymmetry, we will focus on this scenario in the following; it is also worth pointing out that the 2HDM of type I does not face comparable tension as the 2HDM when considered from the perspective of an SFOEWPT~\cite{Basler:2016obg,Atkinson:2021eox,Atkinson:2022pcn,Goncalves:2023svb}.
In the type II case, the Yukawa interactions reduce to
\begin{equation}
\label{eq:2HDM-Yukawa2}
	\mathcal{L}^{\text{dim-4}}_{\text{Yuk}} = - Y^e_{1}\bar{L}\Phi_1 e - Y^d_{1}\bar{Q}\Phi_1 d- Y^u_{2}\bar{Q}\tilde{\Phi}_2 u + \text{h.c.}\,.
\end{equation}
After spontaneous symmetry breaking, we have five physical fields, two
CP-even neutral scalars $H$ and $h$ (ordered in mass to $m_H>m_h$),
one CP-odd scalar $A$ and a charged pair $H^{\pm}$. These fields are
related to the interaction fields through the rotation matrix $R(x)$ as:
\begin{equation}
	\begin{pmatrix}
		H \\ h
	\end{pmatrix}=R(\alpha)
	\begin{pmatrix}
		\zeta_{1} \\ \zeta_{2}
	\end{pmatrix}, \hspace{0.3cm}
	\begin{pmatrix}
	G^{0} \\ A
\end{pmatrix}=R(\beta)
\begin{pmatrix}
	\psi_{1} \\ \psi_{2}
\end{pmatrix},  \hspace{0.3cm}
	\begin{pmatrix}
		G^{\pm} \\ H^{\pm}
	\end{pmatrix}=R(\beta)
	\begin{pmatrix}
		\phi_{1}^{\pm} \\ \phi_{2}^{\pm}
	\end{pmatrix}, 
\end{equation}
with
\begin{equation}
R(x) =	\begin{pmatrix}
		\cos x & \sin x \\
		-\sin x & \cos x
	\end{pmatrix}.
	\end{equation}
The mixing angle $\beta$ is also expressed as:
\begin{equation}
	\tan{\beta}= \frac{v_{2}}{v_{1}}\,,
\end{equation} 
which provides the relation to $v\simeq 246~\text{GeV}$ via
$v_{1}= v \cos{\beta}$ and $v_{2}= v \sin{\beta}$.
The Higgs boson couplings to fermions $f$ in the mass basis fields are given by
\begin{multline}
	\label{eq:yuk}
\mathcal{L}^{\text{dim-4}}_{\text{Yuk}} = 
	- \sum \limits_{f= u,d,\ell} \frac{m_f}{v} \left(\xi_h^f \, \bar{f} f h + \xi_H^ f \, \bar{f} f H - i \xi_A^f  \, \bar{f} \gamma_5 f A \right) \\
	+  \left[ \frac{\sqrt{2} V_{ud}}{v} \bar{u} \left(m_d \, \xi_A^d {\text{P}}_{\text{R}} + m_u \, \xi_A^u {\text{P}}_{\text{L}} \right) d \; H^+ + \frac{\sqrt{2}}{v} m_\ell \, \xi_A^l  (\bar{\nu} {\text{P}}_{\text{R}} \ell)  H^+ + {\text{h.c.}}\right]\,,
\end{multline}
where ${\text{P}}_{\text{L,R}}$ are the left and right chirality
projectors and the coupling modifiers $\xi$ for the type II case are listed in 
Tab.~\ref{tab:coupmod}.
The mass-coupling relations will be modified by the dimension-6 interactions which we detail below.

\begin{table}[!b]
	\centering
	\begin{tabular}{|c|c|c|c|c|c|c|}
		\hline
		\rm Model & $\xi_h^u$ & $\xi_h^{d(e)}$  & $\xi_H^u$ & $\xi_H^{d(e)}$ & $\xi_A^u$ & $\xi_A^{d(e)}$ \\
		\hline
		type II &  $\cos\alpha/\sin\beta$  & $-\sin\alpha/\cos\beta$  & $\sin\alpha/\sin\beta$ &  $\cos\alpha/\cos\beta$ & $\cot\beta$ & $\tan\beta$  \\
		\hline
	\end{tabular}
	\caption{Coupling modifiers $\xi$ for 2HDM type II and up- and down-type quarks.}
	\label{tab:coupmod}
\end{table}


Having set the stage of the renormalisable $d=4$ part of the 2HDM, we
can now turn to its EFT deformation. The extension of these Yukawa
interactions to the effective dimension-6 level results from the
class\footnote{The dimension-6 effective operators for 2HDMEFT are
  classified into 8 classes following the convention of the Warsaw basis given in~\cite{Grzadkowski:2010es}.} $\sim\Psi^2\Phi^3$ which modifies the 2HDM Yukawa Lagrangian~\cite{Crivellin:2016ihg,Karmakar:2017yek,Anisha:2019nzx,Banerjee:2020bym}
\begin{equation}
	{\cal{L}}_{\text{EFT}} = {\cal{L}}_{\text{2HDM}} + \sum_{i} 
	{C_{i}\over \Lambda^2 } O_{i}
	\quad \Longrightarrow \quad
	\mathcal{L}^{\text{EFT}}_{\text{Yuk}}  = \mathcal{L}^{\text{dim-4}}_{\text{Yuk}} + \sum_{i}
	{C_{i}\over \Lambda^2} O_{i} \;. \label{eq:dim6yuk}
\end{equation}
Here, $ O_{i}$ are the dimension-6 operators and $C^{i}$ are the corresponding Wilson Coefficients (WCs). 
For our work, we consider operators dealing with the third generation
fermions i.e. $\tau,t,b$.  The structures of these operators are given
explicitly in Tab.~\ref{tab:2HDM-psi2phi3}. For the type II scenario,
the $\mathbb{Z}_2$ symmetry is enforced with the following
transformations in these operators: for the $\tau$ lepton and the $b$
quark, $\Phi_{1} \rightarrow \Phi_{1}$  and $\Phi_{2} \rightarrow
-\Phi_{2}$ and for the $t$ quark $\Phi_{1} \rightarrow -\Phi_{1}$ and
$\Phi_{2} \rightarrow \Phi_{2}$. The operators violating the
$\mathbb{Z}_2$ symmetry are coloured in magenta. This complete set of
operators modifies the fermion mass terms and the scalar-fermions couplings.

\begin{table}[!t]
	\centering
	\renewcommand{\arraystretch}{1.7}
	{\small\begin{tabular}{|c|c||c|c||c|c|}
			\hline \hline
			\textcolor{purple}{$O_{L\tau}^{1(21)}$}&
			\textcolor{purple}{$(\bar{L}\tau\Phi_1)(\Phi_2^{\dagger}\Phi_1)$}&
			\textcolor{purple}{$O_{L\tau}^{2(22)}$}&
			\textcolor{purple}{$(\bar{L}\tau\Phi_2)(\Phi_2^{\dagger}\Phi_2)$}&
		    \textcolor{purple}{	$O_{L\tau}^{2(11)}$}&        
			\textcolor{purple}{$(\bar{L}\tau\Phi_2)(\Phi_1^{\dagger}\Phi_1)$}\\
			
			\textcolor{purple}{$O_{L\tau}^{1(12)}$}&
			\textcolor{purple}{$(\bar{L}\tau\Phi_1)(\Phi_1^{\dagger}\Phi_2)$}&
			\textcolor{purple}{$O_{Qb}^{1(21)}$}&        
			\textcolor{purple}{$(\bar{Q}\,b\,\Phi_1)(\Phi_2^{\dagger}\Phi_1)$}&
			\textcolor{purple}{$O_{Qb}^{2(22)}$}&
			\textcolor{purple}{$(\bar{Q}\,b\,\Phi_2)(\Phi_2^{\dagger}\Phi_2)$}\\
			
			\textcolor{purple}{$O_{Qb}^{2(11)}$}&
			\textcolor{purple}{$(\bar{Q}\,b\,\Phi_2)(\Phi_1^{\dagger}\Phi_1)$}&
			\textcolor{purple}{$O_{Qb}^{1(12)}$}&
			\textcolor{purple}{$(\bar{Q}\,b\,\Phi_1)(\Phi_1^{\dagger}\Phi_2)$}&
			$O_{Qt}^{2(22)}$&        
			$(\bar{Q}t\tilde{\Phi}_2)(\Phi_2^{\dagger}\Phi_2)$\\
			
			$O_{Qt}^{1(12)}$&
			$(\bar{Q}\,t\,\tilde{\Phi}_1)(\Phi_1^{\dagger}\Phi_2)$&
			$O_{Qt}^{2(11)}$&        
			$(\bar{Q}\,t\,\tilde{\Phi}_2)(\Phi_1^{\dagger}\Phi_1)$&
			$O_{Qt}^{1(21)}$&
			$(\bar{Q} \,t \, \tilde{\Phi}_1)(\Phi_2^{\dagger}\Phi_1)$\\
			
			$O_{L\tau}^{1(11)}$&
			$(\bar{L}\tau\Phi_1)(\Phi_1^{\dagger}\Phi_1)$&
			$O_{L\tau}^{2(12)}$&
		    $(\bar{L}\tau\Phi_2)(\Phi_1^{\dagger}\Phi_2)$&
			$O_{L\tau}^{1(22)}$&
			$(\bar{L}\tau\Phi_1)(\Phi_2^{\dagger}\Phi_2)$\\
			
		    $O_{L\tau}^{2(21)}$&
			$(\bar{L}\tau\Phi_2)(\Phi_2^{\dagger}\Phi_1)$&
			$O_{Qb}^{1(11)}$&
		    $(\bar{Q}\,b\,\Phi_1)(\Phi_1^{\dagger}\Phi_1)$&
			$O_{Qb}^{2(12)}$&
			$(\bar{Q}\,b\,\Phi_2)(\Phi_1^{\dagger}\Phi_2)$\\
			
			$O_{Qb}^{1(22)}$&
			$(\bar{Q}\,b\,\Phi_1)(\Phi_2^{\dagger}\Phi_2)$&
		    $O_{Qb}^{2(21)}$&
			$(\bar{Q}\,b\,\Phi_2)(\Phi_2^{\dagger}\Phi_1)$&
			\textcolor{purple}{$O_{Qt}^{1(11)}$}&
			\textcolor{purple}{$(\bar{Q}\,t\,\tilde{\Phi}_1)(\Phi_1^{\dagger}\Phi_1)$}\\
			
			\textcolor{purple}{$O_{Qt}^{2(21)}$}&
			\textcolor{purple}{$(\bar{Q}\,t\,\tilde{\Phi}_2)(\Phi_2^{\dagger}\Phi_1)$}&        
			\textcolor{purple}{$O_{Qt}^{1(22)}$}&
			\textcolor{purple}{$(\bar{Q}\,t\,\tilde{\Phi}_1)(\Phi_2^{\dagger}\Phi_2)$}&
			\textcolor{purple}{$O_{Qt}^{2(12)}$}&
			\textcolor{purple}{$(\bar{Q}\,t\,\tilde{\Phi}_2)(\Phi_1^{\dagger}\Phi_2)$}\\				
			\hline \hline
	\end{tabular}}
	\caption{Dimension-6 2HDMEFT operators of class
          $\Psi^2\Phi^3$. Each of these operators has a distinct
          Hermitian conjugate.  Here, $\bar{L} = (\bar{\nu}_{\tau}
          \hspace{0.2cm} \bar{\tau})$ and $\bar{Q} = (\bar{t}
          \hspace{0.2cm} \bar{b})$. The operators coloured in magenta
          violate the $\mathbb{Z}_2$ symmetry.}
	\label{tab:2HDM-psi2phi3}
\end{table}  

In the broken phase, these interactions lead to corrections to the fermion masses compared to the dimension-4 mass-Yukawa coupling relation,
\begin{multline}
	\Delta{M}_{\Psi}=  -\frac{1}{2\sqrt{2}\Lambda^2}\big[C_{Q\Psi}^{1(11)} v_1^3+v_1^2v_2(C_{Q\Psi}^{1(12)}+C_{Q\Psi}^{1(21)}+C_{Q\Psi}^{2(11)})\\+v_1v_2^2(C_{Q\Psi}^{1(22)}+C_{Q\Psi}^{2(12)}+C_{Q\Psi}^{2(21)})+C_{Q\Psi}^{2(22)}v_2^3\big]~\text{for}~\Psi\equiv \{t,b,\tau\}.
\end{multline}
For the considered type II scenario, the  modified third-generation fermion mass terms are then
\begin{subequations}
\begin{alignat}{4}
M_{t}=&  \frac{v_{2}}{\sqrt{2}}\Big[Y^{t}_{2}-\frac{1}{2\Lambda^2}\Big(\textcolor{purple}{C_{Qt}^{1(11)}} \frac{v_1^3}{v_2} \nonumber
   +v_1^2(C_{Qt}^{1(12)}+C_{Qt}^{1(21)}+C_{Qt}^{2(11)})\\& \hspace{5cm}\textcolor{purple}{+v_1v_2(C_{Qt}^{1(22)}+C_{Qt}^{2(12)}
   +C_{Qt}^{2(21)})}+C_{Qt}^{2(22)}v_2^2\Big)\Big]\,, \\
M_{b}= & \frac{v_{1}}{\sqrt{2}}\Big[Y^{b}_{1}-\frac{1}{2\Lambda^2}\Big(C_{Qb}^{1(11)} v_1^2\nonumber 
\textcolor{purple}{+v_1 v_{2}(C_{Qb}^{1(12)}+C_{Qb}^{1(21)}+C_{Qb}^{2(11)})}\\&\hspace{5cm}+v_2^2(C_{Qb}^{1(22)}+C_{Qb}^{2(12)}+C_{Qb}^{2(21)})\textcolor{purple}{+C_{Qb}^{2(22)}\frac{v_2^3}{v_1}}\Big)\Big]\,, \\
M_{\tau}= & \frac{v_{1}}{\sqrt{2}}\Big[Y^{\tau}_{1}-\frac{1}{2\Lambda^2}\Big(C_{L\tau}^{1(11)} v_1^2 \textcolor{purple}{+v_1 v_{2}(C_{L\tau}^{1(12)}+C_{L\tau}^{1(21)}+C_{L\tau}^{2(11)})}\nonumber\\&\hspace{5cm}+v_2^2(C_{L\tau}^{1(22)}+C_{L\tau}^{2(12)}+C_{L\tau}^{2(21)})\textcolor{purple}{+C_{L\tau}^{2(22)}\frac{v_2^3}{v_1}}\Big)\Big]\,.
\end{alignat}
\end{subequations}
These mass-coupling relations are different to the ones for $d=4$ quoted in \eqref{eq:yuk}, but we recover $M=m$ for $\Lambda\to\infty$. For our work, the 2HDM dim-4 Yukawa couplings given in Eq.~\eqref{eq:2HDM-Yukawa2} are specifically for the third-generation fermions. Taking masses of fermions as the dimension-6 extended  input quantities, i.e. 
\begin{equation}
	M_{t}= \frac{v_{2}}{\sqrt{2}}\mathcal{Y}^{t}_{2}\,, \hspace{0.5cm} 
	M_{b}= \frac{v_{1}}{\sqrt{2}}\mathcal{Y}^{b}_{1}\,, \hspace{0.5cm} 
	M_{\tau}= \frac{v_{1}}{\sqrt{2}}\mathcal{Y}^{\tau}_{1}\,, 
	\label{eq:redefyuk} 
\end{equation}
the dimension-4 Yukawa couplings are redefined  as
\begin{subequations}
\begin{alignat}{4}
	Y^{t}_{2}  \rightarrow &\;  \mathcal{Y}^{t}_{2} +\frac{1}{2\Lambda^2}\Big(\textcolor{purple}{C_{Qt}^{1(11)}} \frac{v_1^3}{v_2} +v_1^2(C_{Qt}^{1(12)}+C_{Qt}^{1(21)}+C_{Qt}^{2(11)})\\&\hspace{5cm} \textcolor{purple}{+v_1v_2(C_{Qt}^{1(22)}+C_{Qt}^{2(12)}+C_{Qt}^{2(21)})}+C_{Qt}^{2(22)}v_2^2\Big)\,,\nonumber \\
	Y^{b}_{1} \rightarrow &\; \mathcal{Y}^{b}_{1} +\frac{1}{2\Lambda^2}\Big(C_{Qb}^{1(11)} v_1^2 \textcolor{purple}{+v_1 v_{2}(C_{Qb}^{1(12)}+C_{Qb}^{1(21)}+C_{Qb}^{2(11)})}\\ &\hspace{5cm}+v_2^2(C_{Qb}^{1(22)}+C_{Qb}^{2(12)}+C_{Qb}^{2(21)})\textcolor{purple}{+C_{Qb}^{2(22)}\frac{v_2^3}{v_1}}\Big)\,,\nonumber \\
	Y^{\tau}_{1} \rightarrow & \; \mathcal{Y}^{\tau}_{1} +\frac{1}{2\Lambda^2}\Big(C_{L\tau}^{1(11)} v_1^2  \textcolor{purple}{+v_1 v_{2}(C_{L\tau}^{1(12)}+C_{L\tau}^{1(21)}+C_{L\tau}^{2(11)})}\\&\hspace{5cm}+v_2^2(C_{L\tau}^{1(22)}+C_{L\tau}^{2(12)}+C_{L\tau}^{2(21)})\textcolor{purple}{+C_{L\tau}^{2(22)}\frac{v_2^3}{v_1}}\Big)\,. \nonumber  
\end{alignat}
\end{subequations}
These replacements shift the dimension-6-induced coupling modifications into the Higgs-fermion interactions for given fermion masses;
the coupling modifiers $\xi$ mentioned in Tab.~\ref{tab:coupmod} get additional dimension-6 corrections. The modified scalar fermion couplings are given by (assuming $V_{tb}=1$)
\begin{subequations}
\label{eq:coupmods}
\begin{eqnarray}
	\xi^{t}_{h}& = & \frac{\cos{\alpha}}{\sin{\beta}}+  \frac{v^3}{M_t}\frac{1}{\sqrt{2} \Lambda^2}\Big[-  C_{Qt}^{2(22)}\cos{\alpha} \sin^2{\beta} \textcolor{purple}{+\frac{ \cos^2{\beta}}{2}\Big(\frac{\cos{\alpha} \cos{\beta}}{\sin{\beta}} + 3 \sin{\alpha}\Big)C_{Qt}^{1(11)}} \nonumber \\ 
		&& \textcolor{purple}{-\frac{\sin{\beta}}{2}\cos{(\alpha+\beta)}\Big(C_{Qt}^{2(21)}+C_{Qt}^{2(12)}+C_{Qt}^{1(22)} \Big) }+\cos{\beta} \sin{\beta} \sin{\alpha} \; \Big( C_{Qt}^{1(12)} + C_{Qt}^{1(21)} + C_{Qt}^{2(11)}\Big) \Big] \,, \nonumber \\
	\label{eq:higgsmod}\\
	\xi^{t}_{H}& =  & \frac{\sin{\alpha}}{\sin{\beta}}+  \frac{v^3}{M_t}\frac{1}{\sqrt{2} \Lambda^2}\Big[-  C_{Qt}^{2(22)} \sin{\alpha} \sin^2{\beta}  \textcolor{purple}{+\frac{ \cos^2{\beta}}{2} \Big(\frac{\sin{\alpha} \cos{\beta}}{\sin{\beta}} - 3 \cos{\alpha}\Big) C_{Qt}^{1(11)} }\nonumber \\ 
	&&\textcolor{purple}{-\frac{\sin{\alpha}}{2} \sin{(\alpha+\beta)}\Big(C_{Qt}^{2(21)}+C_{Qt}^{2(12)}+C_{Qt}^{1(22)}\Big)} -\cos{\beta} \sin{\beta} \cos{\alpha} \; ( C_{Qt}^{1(12)} + C_{Qt}^{1(21)} + C_{Qt}^{2(11)}) \Big] \,, \nonumber \\
	\label{eq:hhiggsmod}\\
		\xi^{t}_{A}& = & \cot{\beta}+  \frac{v^3}{M_t}\frac{1}{\sqrt{2} \Lambda^2}\Big[ \textcolor{purple}{\frac{ \sin{\beta}}{2}(-C_{Qt}^{2(21)}+C_{Qt}^{2(12)}+C_{Qt}^{1(22)} ) } +\cos{\beta} \; C_{Qt}^{1(12)} \textcolor{purple}{+\frac{\cot{\beta} \cos{\beta}}{2} C_{Qt}^{1(11)} } \Big] \,, \nonumber \\ \label{eq:ahiggsmod}
	\\
	\xi^{b}_{h}& = & -\frac{\sin{\alpha}}{\cos{\beta}}+  \frac{v^3}{M_b}\frac{1}{\sqrt{2} \Lambda^2}\Big[\textcolor{purple}{- C_{Qb}^{2(22)}  \sin^2{\beta} \left( \frac{3 \cos{\alpha}}{2}+ \frac{\sin{\alpha}\sin{\beta}}{2 \cos{\beta}}\right) } +\sin{\alpha} \cos^2{\beta} \; C_{Qb}^{1(11)}  \nonumber \\ &&  \textcolor{purple}{-\frac{\cos{\beta}}{2}\cos{(\alpha+\beta)}\Big( C_{Qb}^{1(12)} + C_{Qb}^{1(21)} + C_{Qb}^{2(11)}\Big) }-\cos{\alpha} \cos{\beta} \sin{\beta}\; \left(C_{Qb}^{2(21)}+C_{Qb}^{2(12)}+C_{Qb}^{1(22)} \right) \Big] \,, \nonumber \\
		\\
	\xi^{b}_{H}& = & \frac{\cos{\alpha}}{\cos{\beta}}+  \frac{v^3}{M_b}\frac{1}{\sqrt{2} \Lambda^2}\Big[ \textcolor{purple}{C_{Qb}^{2(22)}  \sin^2{\beta} \left( -\frac{3 \sin{\alpha}}{2}+ \frac{\cos{\alpha}\sin{\beta}}{2 \cos{\beta}}\right) }-\cos{\alpha} \cos^2{\beta} \; C_{Qb}^{1(11)}  \nonumber \\ &&  \textcolor{purple}{-\frac{\cos{\beta}}{2}\sin{(\alpha+\beta)}\Big( C_{Qb}^{1(12)} + C_{Qb}^{1(21)} + C_{Qb}^{2(11)}\Big)} -\sin{\alpha} \cos{\beta} \sin{\beta} \; \left(C_{Qb}^{2(21)}+C_{Qb}^{2(12)}+C_{Qb}^{1(22)} \right) \Big] \,, \nonumber \\
	\\
	 \xi^{b}_{A}& = & \tan{\beta}+\frac{v^3}{M_b}\frac{1}{\sqrt{2} \Lambda^2}\Big[\textcolor{purple}{ C_{Qb}^{2(22)}  \frac{\tan{\beta}\sin{\beta}}{2} } +  \sin{\beta} \; C_{Qb}^{2(12)}    
	\textcolor{purple}{+ \frac{  \cos{\beta}}{2} \left( C_{Qb}^{1(12)} - C_{Qb}^{1(21)} + C_{Qb}^{2(11)} \right)} \Big] \,,
	\nonumber\\ \\
	\xi^{\tau}_{h}& = & -\frac{\sin{\alpha}}{\cos{\beta}}+  \frac{v^3}{M_\tau}\frac{1}{\sqrt{2} \Lambda^2}\Big[\textcolor{purple}{- C_{L\tau}^{2(22)}  \sin^2{\beta} \left( \frac{3 \cos{\alpha}}{2}+ \frac{\sin{\alpha}\sin{\beta}}{2 \cos{\beta}}\right) } +\sin{\alpha} \cos^2{\beta} \; C_{L\tau}^{1(11)}  \nonumber \\ &&  \textcolor{purple}{-\frac{\cos{\beta}}{2}\cos{(\alpha+\beta)}\Big( C_{L\tau}^{1(12)} + C_{L\tau}^{1(21)} + C_{L\tau}^{2(11)}\Big) }-\cos{\alpha} \cos{\beta} \sin{\beta}\; \left(C_{L\tau}^{2(21)}+C_{L\tau}^{2(12)}+C_{L\tau}^{1(22)} \right) \Big] \,, \nonumber \\
	\\
    \xi^{\tau}_{H}& = & \frac{\cos{\alpha}}{\cos{\beta}}+  \frac{v^3}{M_\tau}\frac{1}{\sqrt{2} \Lambda^2}\Big[ \textcolor{purple}{C_{L\tau}^{2(22)}  \sin^2{\beta} \left( -\frac{3 \sin{\alpha}}{2}+ \frac{\cos{\alpha}\sin{\beta}}{2 \cos{\beta}}\right) }-\cos{\alpha} \cos^2{\beta} \; C_{L\tau}^{1(11)}  \nonumber \\ &&  \textcolor{purple}{-\frac{\cos{\beta}}{2}\sin{(\alpha+\beta)}\Big( C_{L\tau}^{1(12)} + C_{L\tau}^{1(21)} + C_{L\tau}^{2(11)}\Big)}-\sin{\alpha} \cos{\beta} \sin{\beta} \; \left(C_{L\tau}^{2(21)}+C_{L\tau}^{2(12)}+C_{L\tau}^{1(22)} \right) \Big] \,,\nonumber 
    \\ \\
	 \xi^{\tau}_{A}& = & \tan{\beta}+\frac{v^3}{M_\tau}\frac{1}{\sqrt{2} \Lambda^2}\Big[\textcolor{purple}{   \frac{\sin{\beta}\tan{\beta}}{2} C_{L\tau}^{2(22)} } +  \sin{\beta} \; C_{L\tau}^{2(12)}  
	 	\textcolor{purple}{+ \frac{  \cos{\beta}}{2} \left( C_{L\tau}^{1(12)} - C_{L\tau}^{1(21)} + C_{L\tau}^{2(11)}\right)} \Big], \nonumber \\
\end{eqnarray}
\end{subequations}
which reduce to the usual 2HDM relations when decoupling $\Lambda\to \infty$.

\section{Effective Potential at Finite Temperature}
\label{sec:EPT}
The Yukawa interactions detailed above are joined by the renormalisable (dimension 4) Higgs potential~\cite{Gunion:1989we,Gunion:2002zf} 
\begin{multline}
\label{eq:2HDMsc}
 V_{\text{d4}}(\Phi_1, \Phi_2) = {m}^2_{11}(\Phi_1^{\dagger}\Phi_1) + {m}^2_{22}(\Phi_2^{\dagger}\Phi_2) - {m}^2_{12}(\Phi_1^{\dagger}\Phi_2 + \Phi_2^{\dagger}\Phi_1) + {{\lambda}_1}(\Phi_1^{\dagger}\Phi_1)^2 + {{\lambda}_2}(\Phi_2^{\dagger}\Phi_2)^2  \\
 +  {\lambda}_3(\Phi_1^{\dagger}\Phi_1)(\Phi_2^{\dagger}\Phi_2) + {\lambda}_4(\Phi_1^{\dagger}\Phi_2)(\Phi_2^{\dagger}\Phi_1) +
\frac{1}{2}{\lambda}_5[(\Phi_1^{\dagger}\Phi_2)^2 + (\Phi_2^{\dagger}\Phi_1)^2]  \\
+  \left({\lambda}_6(\Phi_1^{\dagger}\Phi_1) +{\lambda}_7(\Phi_2^{\dagger}\Phi_2)\right)\left(\Phi_1^{\dagger}\Phi_2 + \Phi_2^{\dagger}\Phi_1\right)\,,
\end{multline}
and we will focus on the CP-even case, $\lambda_6=\lambda_7=0$. Furthermore, we will only consider the soft $\mathbb{Z}_2$ breaking terms $\sim m_{12}^2$, ignoring the magenta couplings detailed above for the Yukawa interactions. We will limit our discussion to the top quark-specific interactions in the following.

The analysis of the symmetry properties at finite temperatures, see e.g.~\cite{Quiros:1994dr}, requires the calculation of the one-loop effective (Coleman-Weinberg) potential at zero temperature~\cite{Coleman:1973jx}
in addition to temperature corrections and associated Daisy resummation~\cite{Dolan:1973qd,Carrington:1991hz,Quiros:1999jp}. The potential is most economically calculated as outlined
in~\cite{Dolan:1973qd,Jackiw:1974cv}, yielding
\begin{equation}
    V_\text{eff}^{(1)} (\vec{\omega},T) =\sum_{X=S,G,F} (-1)^{2s_X}
  (1+2s_X) I^X \,,
\label{eq:effpot}
\end{equation}
for a general vacuum configuration $\vec{\omega}$ in the scalar $SU(2)$ space of Eq.~\eqref{eq:higgsdoublets}. Equation~\eqref{eq:effpot} sums over
scalar ($S$), gauge field ($G$), fermion ($F$) contributions with spin quantum numbers $s_{S,G,F}=0,1,1/2$ and associated one loop contributions
\begin{equation}
    \label{eq:1LoopIntsFT}
\begin{split}
        I^S &= \frac{T}{2}\sum_n^\text{Bos}\int\frac{d^3 k}{(2\pi)^3} \sum_i \left[\log \det\left(-\mathcal{D}^{-1}_{S,\, i}\right)\right]\,,\\
        I^{G} &= \frac{T}{2}\sum_n^\text{Bos}\int\frac{d^3 k}{(2\pi)^3} \sum_i \left[\log \det\left(-\mathcal{D}^{-1}_{GB,\, i}\right)\right]\,,\\
        I^F &= -T \sum_n^\text{Ferm}\int\frac{d^3 k}{(2\pi)^3} \sum_i \left[\log \det\left(-\mathcal{D}^{-1}_{F,\, i}\right)\right]\,,
\end{split}
\end{equation}
where we have introduced the corresponding inverse propagators as ${\cal{D}}^{-1}_X$. At finite temperature the periodicity conditions on the two-point function
require us to sum over the discrete Matsubara modes~\cite{Matsubara:1955ws} in momentum space, e.g. ${\cal{D}}^{-1}_S = \omega_n^2+\omega_k^2$ with
\newcommand{\MSbar}{$\overline{\text{MS}}$}
\newcommand{\vvec}{\bm}
\begin{equation}
\begin{split}
    \omega_n^2 &= (2n\pi T)^2\,,\quad n\in\mathbb{N}_0 \\
    \omega_k^2 &= \vvec{k}^2+m^2\,,
\end{split}
\end{equation}
in the imaginary time formalism.
The integrals of Eq.~\eqref{eq:1LoopIntsFT} can be evaluated in the \MSbar~scheme
\begin{equation}
\begin{split}
    I^X_{\overline{MS}} &= \frac{m_X^4}{64 \pi^2}\left[\log\left(\frac{m_X^2}{\mu^2}\right)-k_X\right] + \frac{T^4}{2\pi^2} J_\pm\left(\frac{m_X^2}{T^2}\right)\\
   & = V^X_\text{CW}(\Phi_1,\Phi_2) +  V^X_T(\Phi_1,\Phi_2)\,,
\end{split}
\end{equation}
which shows the factorisation into the temperature-independent Coleman-Weinberg (CW) contribution and a temperature-dependent contribution. The ultraviolet (UV) finite constants $k_X$
are given by
\begin{equation}
    k_X=\begin{cases}
        \frac{5}{6},\quad X=G\\
        \frac{3}{2},\quad X=S,F
    \end{cases}
\end{equation}
and the thermal fermionic $(+)$ and bosonic $(-)$ function $J_\pm$~\cite{Dolan:1973qd,Quiros:1994dr,Quiros:1999jp}
\begin{align}
    J_{\pm}(x^2) = \int_0^\infty \text{d}k \, k^2 \log\left(1 \pm e^{-\sqrt{k^2+x^2}}\right)\,.
\end{align}

The presence of Matsubara zero-modes leads to infrared problems linked to the breakdown of perturbation theory at high temperatures~\cite{Weinberg:1974hy}. These infrared problems
are resolved through reordering the perturbative series expansion by including thermal corrections $\Pi$ to the masses, which re-sums the problematic direction of the expansion parameters~\cite{Carrington:1991hz,Parwani:1991gq,Arnold:1992fb,Kapusta:2006pm,Arnold:1992rz,Quiros:1994dr}. Concretely, we employ the Arnold-Espinosa approach~\cite{Arnold:1992rz}
replacing
\begin{equation}
V_T \to V_T  + V_{\text{daisy}} \,,
\end{equation}
with
\begin{equation}
V_{\text{daisy}}= - \frac{T}{12\pi} \left[
  \sum_{i=1}^{n_{\text{Higgs}}} \left( (\overline{m}_i^2)^{3/2} -
    (m_i^2)^{3/2} \right) + \sum_{a}^{n_{\text{gauge}}} \left(
    (\overline{m}_a^2)^{3/2} - (m_a^2)^{3/2} \right) \right] \,.
\end{equation}
The $\overline m$ masses are derived by including thermal mass corrections in the hard thermal limit. In total, the relevant 1-loop potential for our study is given by
\begin{equation}
V(T) = V_{\text{d4}}(T=0) + V_\text{CW}(T=0) + V_T(T) + V_{\text{daisy}}(T)\,.
\end{equation}
The modifications of the Yukawa couplings (together with correlated four- and five-point interactions) outlined in the previous section lead to a modification of the contributions of $I^F$ through new $m_F(\Phi_{1,2})$ contributions, thus changing the $V(T)$ away from its expectation in the $d=4$ 2HDM at $T=0$. These changes are mirrored in the temperature-dependent part alongside modifications to the plasma interactions parametrised by $\overline m$: Effective interactions will typically introduce new contributions to the thermal masses~\cite{Bodeker:2004ws,Croon:2020cgk}, which we have included throughout (it is worth highlighting though that these do not play a relevant role for the parameter choices that we consider in this work).\footnote{Note also that the redefinition of the Yukawa interactions of Eq.~\eqref{eq:redefyuk} already changes the dependence $M_F(\Phi_1,\Phi_2)$ such that only in the vacuum at $T=0$ we recover the effective dimension-4 relations.}

\begin{table}[t!]
\centering
\renewcommand{\arraystretch}{1.8}
\begin{tabular}{|c|c||c|c|}
\hline \hline
$O_6^{111111}$& $(\Phi_1^{\dagger}\Phi_1)^3$ & $O_6^{222222}$ & $(\Phi_2^{\dagger}\Phi_2)^3$ \\
$O_6^{111122}$& $(\Phi_1^{\dagger}\Phi_1)^2(\Phi_2^{\dagger}\Phi_2)$& $O_6^{112222}$& $(\Phi_1^{\dagger}\Phi_1)(\Phi_2^{\dagger}\Phi_2)^2$ \\
$O_6^{122111}$& $(\Phi_1^{\dagger}\Phi_2)(\Phi_2^{\dagger}\Phi_1)(\Phi_1^{\dagger}\Phi_1)$& $O_6^{122122}$& $(\Phi_1^{\dagger}\Phi_2)(\Phi_2^{\dagger}\Phi_1)(\Phi_2^{\dagger}\Phi_2)$\\
$O_6^{121211}$&$(\Phi_1^{\dagger}\Phi_2)^2(\Phi_1^{\dagger}\Phi_1)$ + h.c. &$O_6^{121222}$&$(\Phi_1^{\dagger}\Phi_2)^2(\Phi_2^{\dagger}\Phi_2)$ + h.c. \\\hline \hline
\end{tabular}
\caption{Dimension-6 operators of class $\Phi^6$  involving  $\Phi_1$ and $\Phi_2$.}\label{tab:phi6op}
\end{table}   

As done in Refs.~\cite{Basler:2019iuu,Basler:2021kgq}, it is convenient to mirror on-shell renormalisation conditions by considering additional finite counter-term contributions at $T=0$ to enforce an agreement between tree-level and one-loop effective potential minima, masses and mixing which is expressed by
\begin{equation}
\label{eq:cts}
0 = {\partial \over \partial {\phi_i}} 
(V^{\text{CW}}+V^{\text{CT}})\bigg|_{\vec{\bar{\omega}}_{\text{tree}}}
= {\partial ^2\over \partial {\phi_i}\partial{\phi_i}}  (V^{\text{CW}}+V^{\text{CT}})\bigg|_{\vec{\bar{\omega}}_{\text{tree}}}.
\end{equation}
Here we denote $\phi_i$ as the degrees of freedom in Eq.~\eqref{eq:higgsdoublets} and have further defined $\vec{\bar{\omega}}_{\text{tree}}$ as the vacuum selected by Eq.~\eqref{eq:2HDMsc}, which can be aligned in our CP-even case without loss of generality in the $(\zeta_1,\zeta_2)$ direction.

\begin{figure}[!t]
\centering
\includegraphics[width=0.5\textwidth]{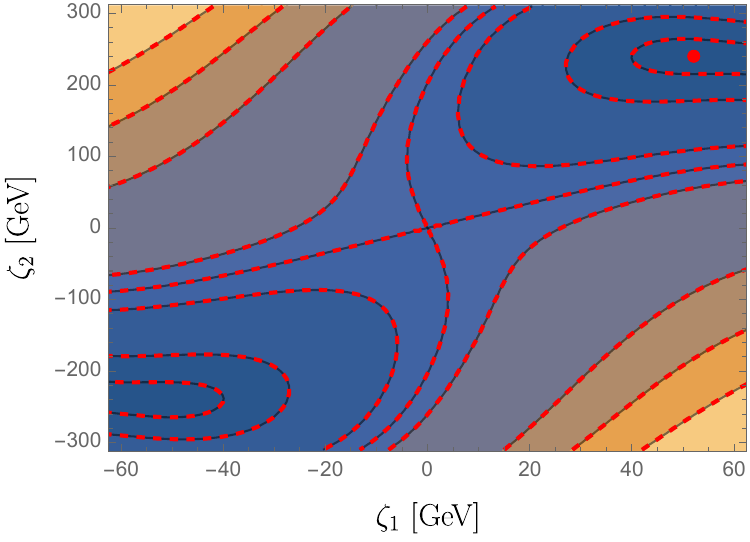}
\caption{The 1-loop $T=0$ vacuum structure (in arbitrary units) in the $\zeta_1,\zeta_2$ parameter space for a representative parameter choice 
$\lambda_1 = 2.74, \lambda_2= 0.24, \lambda_3 = 5.53, \lambda_4= -2.59, \lambda_5 = -2.23, (m_{11}^2,m_{22}^2,m_{12}^2)=
 (11212.6, -6324.6 , 7738.6)~\text{GeV}^2$. This 
gives rise to the tree-level vacuum depicted
by the red dot. The solid contours are computed from the top quark contribution to the effective potential linearised in $\Lambda$, which admits a solution to Eqs.~\eqref{eq:cts}. Overlayed in dashed
contours is the full, un-truncated fermionic contribution to the effective potential employing the solution in the linearised approximation. The Wilson coefficients are chosen 
$C^{2(22)}_{Qt} = C^{1(12)}_{Qt} = C^{2(11)}_{Qt} = C^{1(22)}_{Qt} = C^{1(21)}_{Qt} = 4\pi$, $\Lambda=1~\text{TeV}$, indicating that the linearised approximation is under good control for up to relatively large, yet perturbative Wilson coefficient choices.\label{fig:vacstruc}}
\end{figure}

When considering effective field theories, these requirements are subtle. The CW effective potential re-sums the dimension-6 EFT insertions to all orders in the $\Lambda^{-2}$ expansion. In general, this means that the system of equations, Eq.~\eqref{eq:cts}, is over-constrained when only considering the renormalisation of the $d=4$ parameters. But also including the scalar $d=6$ interactions of Tab.~\ref{tab:phi6op} (see also~\cite{Anisha:2022hgv,Anisha:2019nzx}) are insufficient unless the
effective potential is truncated at $d=6$. For investigations using numerical implementations such as {\tt{BSMPT}}~\cite{Basler:2018cwe,Basler:2020nrq,BSMPTv3}, this poses a technical difficulty as the expansion in $\Lambda$ is no longer under analytical control: Eq.~\eqref{eq:cts} are unattainable for general parameter choices. Analytical cross-checks show that this, however, does not lead to numerically relevant deviations for perturbative Wilson coefficient choices where we can trust our results in the first place. This is demonstrated in Fig.~\ref{fig:vacstruc} which shows the top-quark contribution to the effective potential in the presence of top-specific Wilson coefficients.

\section{Phenomenology of the Electroweak Phase Transition}

\subsection{Scan Methodology}
\label{sec:scan}
The exploration of the top-EFT extended 2HDM is performed numerically
using {\tt
  ScannerS}~\cite{Coimbra:2013qq,ScannerS,Muhlleitner:2020wwk}. We have
modified the original implementation of the real 2HDM (\texttt{R2HDM}) to include the operators given in Tab.~\ref{tab:2HDM-psi2phi3}. Furthermore, we modified accordingly the code \texttt{HDECAY}~\cite{Djouadi:1997yw, Harlander:2013qxa, Djouadi:2018xqq} for the computation of the QCD corrected branching ratios of all scalar particles.
We choose the 2HDM mass spectrum, $\tan\beta$, the soft-breaking $m_{12}^2$, the coupling of the heavy CP-even Higgs boson to massive gauge bosons $c_{HVV}$, as well as the corresponding Wilson coefficients (setting $\Lambda=1$ TeV) as input parameters. The light Higgs boson is selected to have a mass of
\begin{equation}
m_h = 125.09~\text{GeV}
\end{equation}
and behave SM-like. Points are generated from random numbers and their consistency with phenomenological
constraints is checked by {\tt{ScannerS}} using 
{\tt{HiggsBounds}}~\cite{Bechtle:2008jh,Bechtle:2011sb,Bechtle:2013wla,Bechtle:2020pkv} and {\tt{HiggsSignals}}~\cite{Bechtle:2013xfa,Bechtle:2020uwn}.
Flavour constraints are taken into account through consistency with $\mathcal{R}_b$ \cite{Haber:1999zh,Deschamps:2009rh} and
$B\rightarrow X_s \gamma $
\cite{Deschamps:2009rh,Mahmoudi:2009zx,Hermann:2012fc,Misiak:2015xwa,Misiak:2017bgg, Misiak:2020vlo}. 
Given the Yukawa type II considered here, the charged Higgs mass is constrained to be 
$m_{H^\pm} \ge 800$~GeV~\cite{Misiak:2020vlo}, virtually independent of $\tan\beta$.\footnote{The bound on $m_{H^\pm}$ is currently subject to investigation given the recent results by Belle-II~\cite{Belle-II:2022hys,Misiak:2020vlo,priv}.}
The input parameters' minimum and maximum allowed ranges
are provided in Tab.~\ref{table:r2hdmrange}. The imposed limit on the charged Higgs mass effectively removes phenomenological sensitivity to this state, and we will focus on modifications of the neutral states, which are much more accessible at the LHC. The coupling modifiers analogous to Eq.~\eqref{eq:coupmods} are given in appendix \ref{app:coup} for completeness.

\renewcommand{\arraystretch}{1}
\begin{table}[t!]
\begin{center}
\begin{tabular}{|c|c|c|c|c|c|c|}
\hline
$m_{h}$ [GeV] & $m_{H}$ [GeV] & $m_A$ [GeV] & $m_{H^\pm}$ [GeV] & $\tan\beta$ &
$c_{H VV}$ & $m_{12}^2$ [GeV$^2$]\\ \hline \hline
125.09 & 130...3000 & 30...3000 & 800...3000 & 0.8...30 & -0.3...1.0 &
 $10^{-5}$...$10^7$\\ \hline
\end{tabular}
\end{center}
\vspace*{-0.3cm}
\caption{Scan ranges of the 2HDM input parameters. \label{table:r2hdmrange} }
\end{table}
\renewcommand{\arraystretch}{1} 

The experimentally and theoretically validated parameter points found with {\tt ScannerS} are further investigated with our code {\tt BSMPT} \cite{Basler:2018cwe,Basler:2020nrq,BSMPTv3}.
Our scan methodology works like follows:
\begin{enumerate}
    \item We scan for a dim-4 point (all Wilson coefficients
      $C^i_{Qt}=0$) that is in agreement with theoretical and
      experimental constraints with {\tt ScannerS} and whose strength of
      the electroweak phase transition is  $\xi_c^\text{d4}<1$ (no SFOEWPT yet) which we check with {\tt BSMPT}.\\[1em]
        For each dim-6 Wilson coefficient direction $C_{Qt}^i$ we then check the following:
    \item The selected dim-6 direction is varied with $C_{Qt}^i=\pm 0.01$ and we evaluate the response in $\xi_c^\text{d6}$ by tracing the phases in a range of $T_c^\text{d4} \pm \SI{15}{GeV}$ around the dim-4 critical temperature $T_c^\text{d4}$.\footnote{Due to our lack of analytical control over the $\Lambda^{-1}$ expansion, our non-linearized calculation manifests itself into small deviations from the EW minimum at $T=\SI{0}{GeV}$, as well as small deviations from EW symmetry restoration at $T=\SI{300}{GeV}$.
        Because we analytically found them to be numerically irrelevant for perturbative Wilson coefficient choices, we are only interested in studying the impact of $C_{Qt}^i$ on the behaviour of the false and true coexisting minima phases around the dim-4 critical temperature.}
        For the minimum phase tracing, we use the new minimum tracing algorithms of {\tt BSMPTv3} \cite{BSMPTv3}.
    \item From the results for $\xi^\text{d6}_{c,\pm 0.01}$, we make a
      prediction for the Wilson coefficient
      $C_{Qt}^{i,\text{SFOEWPT}}$ leading to an SFOEWPT, assuming a linear response.
    \item The prediction is checked with {\tt ScannerS} including special focus on the $h_{125}t\bar{t}$ coupling. If the predicted point is found to be valid, we derive $\xi_c^\text{d6}$ with {\tt BSMPTv3} as described in step 2. Here, we adjust the temperature ranges for minima tracing iteratively.
    \item We keep the point as a valid linear response dim-6 SFOEWPT point if $\xi_c^\text{d6, pred}$ differs from $\xi_c^\text{SFOEWPT} = 1$ by less than \SI{1}{\%}. For relative differences above \SI{10}{\%} we discard the point due to showing a non-linear response that violates our assumption of perturbative Wilson coefficient choices. 
    For $\SI{1}{\%}<|1-\xi_c^\text{d6, pred}|<\SI{10}{\%}$ we make an updated linearised prediction for the dim-6 Wilson coefficient strength needed for an SFOEWPT based on the previous iteration 
        \begin{align}
            C_{Qt}^{i,\text{SFOEWPT}} = C_{Qt}^{i,\text{prev}} \cdot \left(\frac{1 - \xi_c^\text{d4}}{\xi_c^\text{d6, prev} - \xi_c^\text{d4}}\right)
        \end{align}
    and repeat steps 4. and 5. until a valid linear-response dim-6 SFOEWPT point is found or the point has to be discarded due to detected non-linearities.
\end{enumerate}

\begin{table}[!t]
  \begin{center}
      \begin{tabular}{|c|c|c|c|c|c|c|}
          \hline
          $m_{h}$ [GeV] & $m_{H}$ [GeV] & $m_A$ [GeV] & $m_{H^\pm}$ [GeV] & $\tan\beta$ &
          $c_{H VV}$ & $m_{12}^2$ [GeV$^2$] \\ \hline \hline
          125.09 & 683 & 872 & 868 & 1.658 & 0.00350 &
          205007 \\ \hline
      \end{tabular}\\[0.2cm]
      \begin{tabular}{|c|c|c|}
          \hline
          $T_c^\text{d4}$ [GeV] & $v(T_c)^\text{d4}$ [GeV] & $\xi_c^\text{d4}$\\ \hline \hline
          226.29 & 215.69 & 0.95 \\ \hline
      \end{tabular}
  \end{center}
\vspace*{-0.3cm}
   \caption{\label{tab:one_point_detail}Input parameters of the benchmark point used for Fig.~\ref{fig:one_point_detail}.}
\end{table}
\begin{figure}[!t]
    \centering
    \includegraphics[width=\textwidth]{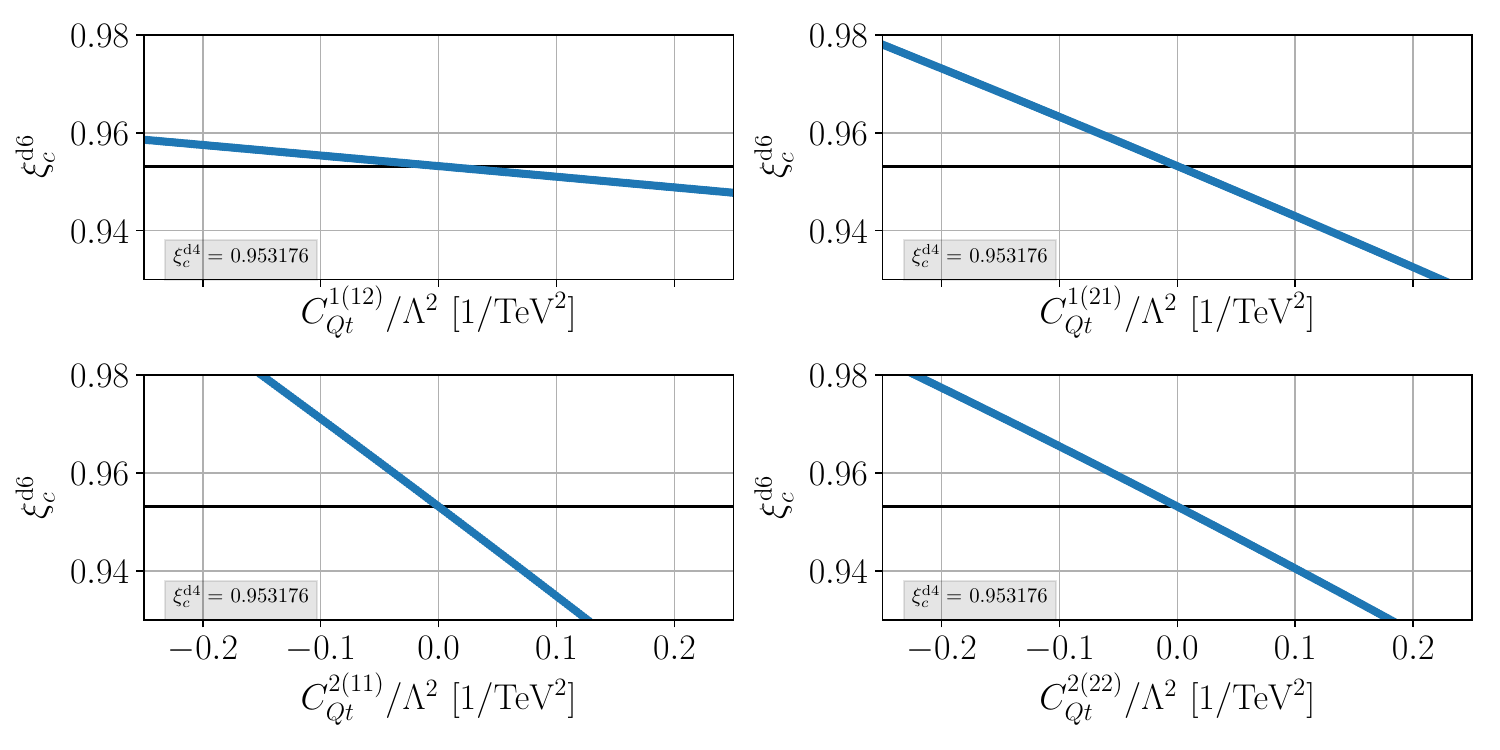}
\vspace*{-0.6cm}
    \caption{Response in $\xi_c^\text{d6}$ (in blue) in all Wilson
      coefficient directions separately for a representative parameter
      point of the Type-2 2HDM with
      $\xi_c^\text{d4}\simeq\num{0.95}$. The dim-4 $\xi_c^\text{d4}$ is
      marked as a black line. The displayed point is given in
      Table~\ref{tab:one_point_detail}.}\label{fig:one_point_detail}
\end{figure}

In Fig.~\ref{fig:one_point_detail} we show the detailed response in $\xi_c^\text{d6}$ for one picked exemplary linear-response parameter point. The point is given in Tab.~\ref{tab:one_point_detail} in detail. As can be seen in these plots the generic response to EFT parameter modifications is linear. This means that although the potential receives non-linear contributions from the EFT correlation changes, these formally higher-order modifications are
not relevant within the region that we study in this work. These results can therefore be taken as a consistency check of the dimension 6 approach outlined in the previous section.

\subsection{Results and Implications of a Top-Philic SFOEWPT}
\label{sec:results}
Small modifications of the top interactions can have a sizeable impact on Higgs signal strengths
\begin{equation}
\mu (h\to XY) = { [\sigma(h) \times \text{BR}(h \to XY)]^\text{d6} \over[\sigma(h) \times \text{BR}(h \to XY)]^\text{d4} }
\end{equation}
where $\sigma(h)$ is the light Higgs production cross section and $\text{BR}(h \to XY)$ the branching ratio into the final state $h\to XY$. In particular,
the branching ratio of the $h\to \gamma \gamma$ decay, which is already significantly constrained
and will provide a formidable avenue to constrain this direction in the future, limits the freedom of BSM interactions. The coupling modifier
of the 2HDM can move quickly away as a function of the Wilson
coefficients from the alignment limit that is preferred by the increasingly SM-consistent
outcome of Higgs measurements at the LHC. This becomes particularly clear in a dedicated scan of {\emph{individual}}
operator directions of Tab.~\ref{tab:2HDM-psi2phi3}. Indeed we find the Higgs signal strength constraints
that are part of our workflow, Sec.~\ref{sec:scan}, limit our freedom of
Wilson coefficients,  
highlighting scan points that achieve $\xi^{\text{d6}}_c>1$ from distances $1-\xi^{\text{d4}}_c\simeq 10\%$, i.e.
we can only bridge small $\text{d4}$ phase transition distances without violating signal strength constraints. This consistency with the SM outcome
naturally moves us to a parameter domain where EFT modifications can be trusted.

\begin{figure}[!t]
\centering
\includegraphics[width=7.5cm]{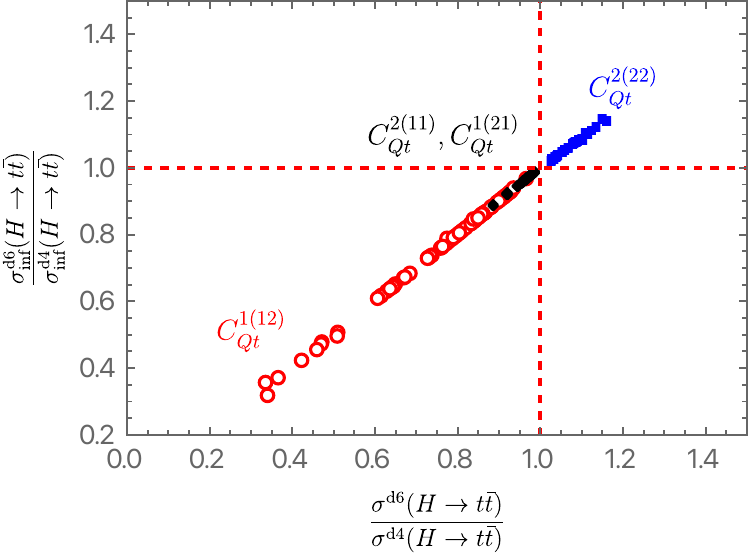}
\hspace{1cm}
\includegraphics[width=7.5cm]{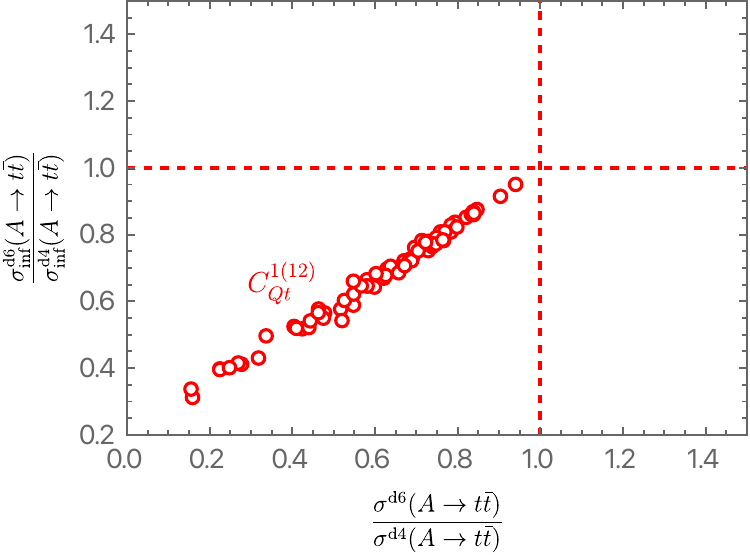}
\hspace{1cm}
\vspace*{-0.6cm}
\caption{\label{fig:ttscan} Correlation of dimension-6 modified signal
  cross sections $\sigma^{\text{d6}}$ relative to their dimension-4 2HDM
  expectation $\sigma^{\text{d4}}$.  The cross sections
  $\sigma^{\text{d4}}_{\text{inf}}, \sigma^{\text{d6}}_{\text{inf}}$
  include interference effects with other signal contributions (e.g. propagating $A,h$ contributions in case of $H$ production) in the
  2HDM as well as, most importantly, interference effects with
  QCD-induced $t\bar t$ production. We include points that are characterised by $\xi^{\text{d4}}_c<0.96$. 
}
\end{figure}

In parallel, we require a priori significant Yukawa-sector modifications to enable a stronger phase transition in comparison with the SM alone (see also the discussion in Ref.~\cite{Baldes:2016rqn} in 
the context of a different model).\footnote{We note at this point that requiring $\xi^{\text{d6}}_c=1$ as a numerical value does not automatically guarantee an SFOEWPT. What we are interested in predominantly in the following are the phenomenological consequences at the LHC that are implied by ``gradients'' in $\xi^{\text{d4}}_c  \to \xi^{\text{d6}}_c\gtrsim 1$. This enables us to qualitatively understand exclusion constraints or the lack of 
new physics signatures through the lens of overcoming the shortfalls of the 2HDM type II.}
Parametric freedom in signal strength constraints (that are included in our scan as described in Sec.~\ref{sec:scan}) can be achieved for mixing angles that reduce the sensitivity to a particular Wilson coefficient for the 125 GeV Higgs.\footnote{The operator $\sim C^{1(21)}_{Qt}$ is particularly worth
highlighting here as it is the only $\mathbb{Z}_2$ symmetry conserving operator that modifies the interactions with the CP-odd scalar, thus offering additional
phenomenological handles at the LHC.} Compatibility of the 2HDM
predictions with the currently observed consistency of $h=H_{\text{SM}}$
favours regions of $\tan\beta \sim {\cal{O}}(1)$ at a coupling modifier
$\xi_h^t\simeq 1$, Eq.~\eqref{eq:higgsmod}, isolating the alignment
limit $\cos\alpha\simeq \sin\beta$ in the Yukawa sector. For these
parameter choices, $\sin\alpha <0$ in our conventions, so that the
negative Wilson coefficient choices that drive  $\xi^{\text{d6}}_c\to 1$
are correlated with slightly enhanced coupling modifiers $\xi^t_h$. This
increases the thermal contribution to the Higgs potential from the
lightest and therefore most relevant degrees of freedom for the thermodynamical problem at hand. 

In case of the $O^{1(12)}_{Qt}$ interactions, the correlated modification for the
heavy $H$, Eq.~\eqref{eq:hhiggsmod}, is then a reduced coupling strength
$|\xi_H^t(C^{1(12)}_{Qt})|<|\xi_H^t(C^{1(12)}_{Qt}=0)|$, which is mirrored by the CP-odd state as
$\tan\beta>0$, Eq.~\eqref{eq:ahiggsmod}. 
The new physics operators can modify the $h$ coupling strengths at the order of 10\% given current Higgs constraints, and a sizeable Wilson coefficient to drive the EWPT requires suppression to maintain consistency with light Higgs observations. 
Depending on the particular regions of parameter space where
$\xi^{\text{d6}}_c\simeq 1$ and agreement with available data can be
achieved in the scan detailed above, an interesting phenomenological
implication arises, which especially highlights
$O^{1(12)}_{Qt}$. Heavy physics parametrised by $C^{1(12)}_{Qt}$
that points towards an SFOEWPT is
correlated with an underproduction of the additional Higgs bosons in the
2HDM in the dominant gluon fusion channels $gg\to H/A\to t\bar t$,
Fig.~\ref{fig:ttscan}. 
The relative reduction due to angular suppression of the Wilson
coefficient to maintain consistency with $h$ data is not given for the
heavy states whose phenomenology therefore significantly departs from
the ${\text{d}}4$ 2HDM expectation. 

Contrary to the $O^{1(12)}_{Qt}$, the structure of $O^{2(22)}_{Qt}$ is such that
\begin{align}
\frac{\xi^{t,\text{d6}}_{H}}{\xi^{t,\text{d6}}_{h}}\bigg|_{O^{2(22)}_{Qt}}&=\frac{\xi^{t,\text{d4}}_{H}}{\xi^{t,\text{d4}}_{h}}\bigg|_{O^{2(22)}_{Qt}}=\tan\alpha\,,\\
\frac{\xi^{t,\text{d6}}_{h}}{\xi^{t,\text{d4}}_{h}}\bigg|_{O^{2(22)}_{Qt}}&= \frac{\xi^{t,\text{d6}}_{H}}{\xi^{t,\text{d4}}_{H}}\bigg|_{O^{2(22)}_{Qt}}= 1 -  C_{Qt}^{2(22)} \frac{v^2}{\Lambda^2}\frac{v}{ \sqrt{2} M_t} \sin^3\beta\,.
\end{align}
Due to the vacuum structure of this operator, the $h$, and $H$ phenomenology modifications are fully correlated, independent of the size of the Wilson coefficient. An enhanced strength of the phase transition then manifests itself through a dedicated pattern in strengths of $H$ vs $h$ interactions that can depart from the 2HDM d4 expectation at 20\% enhancement whilst the CP odd Higgs boson interactions are unchanged to leading approximation.

\begin{figure}[!t]
	\centering
	\includegraphics[width=9.5cm]{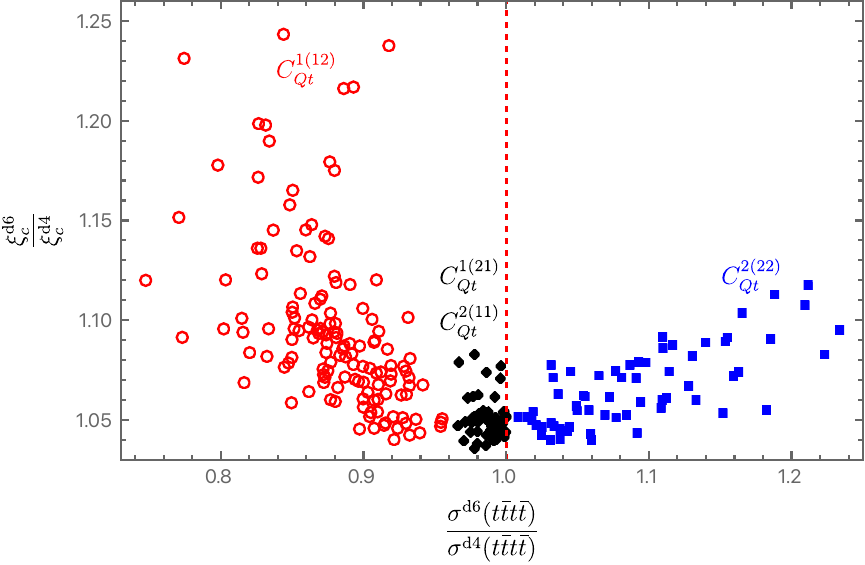}
	\caption{Correlation of the modified production cross-section for $pp \rightarrow t \bar{t}t\bar{t}$ 
		($\sigma^{d6}/\sigma^{d4}$) with modification in the strength of the phase transition ($\xi^{\text{d6}}_{c}/\xi^{\text{d4}}_{c}$). The corresponding contributions for the different effective operators are shown and the points are chosen with $\xi^{\text{d4}}_c<0.96$. \label{fig:ttttscan}}
\end{figure}

Interactions related to $O^{1(21)}_{Qt}, {O}^{2(11)}_{Qt}$
impact the neutral Higgs sector identically, and therefore the
phenomenology is correlated. We, therefore, show their results combined
in Fig.~\ref{fig:ttscan}. The qualitative picture is similar to
$O^{1(12)}_{Qt}$, however, as $\xi^{\text{d6}}_c$ is more
sensitive to the Wilson coefficient in this case. This can be seen, e.g., in the steeper gradient displayed for
$C^{1(21)}_{Qt},C^{2(11)}_{Qt}$ for the sample parameter point in Fig.~\ref{fig:one_point_detail} compared to $C^{1(12)}_{Qt}$.
Hence, the quantitative impact is reduced.

It is well-known that these searches are limited by accidental signal
background interference~\cite{Gaemers:1984sj}, however, the reduction in
signal rate does not qualitatively change the observed outcome on which
the model-dependent investigations at the
LHC~(e.g. \cite{CMS:2019pzc,ATLAS:2017snw}) are based. This means that
when considering the correlation changes anticipated from dimension-6
deformations of the 2HDM type II, the established experimental
strategies remain valid. In Fig.~\ref{fig:ttttscan}, we also show the
implications for four top final states $pp\to t\bar t t\bar t$ (which
includes all Higgs contributions in $s$ and $t$ channels). This process
has been motivated as an additional (interference-robust) tool to
constrain or observe new
physics~\cite{Alvarez:2016nrz,Alvarez:2019uxp,Kanemura:2015nza,Blekman:2022jag,Anisha:2023xmh}
(see also the recent LHC results
of~\cite{ATLAS:2023ajo,CMS:2023ftu}). The implications for the four top
final states are identical to $gg\to H/A\to t\bar t$, cf. Fig.~\ref{fig:ttscan}.

What is perhaps most important at this point in the LHC programme is
that when we consider the aforementioned correlation changes that
address cosmological shortcomings of the 2HDM at face value, the LHC
sensitivity is currently {\emph{overestimated}}, predominantly for $C^{1(12)}_{Qt}$, for which also the CP-odd scalar
has a suppressed phenomenology (such states are abundantly produced compared to the CP-even scalar due to a different threshold behavior~\cite{Djouadi:2005gi}).
This alludes to the tantalising possibility that the 2HDM type-II could
indeed be realised at the TeV scale with additional heavier physics
modifying the expected correlations in such a way that the current
constraints are weakened, yet shortfalls of the SM (and the 2HDM) are
cured. This constitutes an exciting prospect for the LHC
Run-3.

\section{Conclusions}
\label{sec:conc}
The requirement of a strong first-order electroweak phase transition is a strong hint for a source of new physics beyond the Standard Model.
Yet, current analyses at the high-energy regime of the LHC seem to indicate that electroweak symmetry breaking is well-described
by the ad-hoc implementation of the SM. On the one hand, these recent observations imply mounting pressure on BSM scenarios such as the 2HDM type II that
we have considered in this work. On the other hand, consistency with the SM hypothesis could indicate top-philic cancellations as part of high-scale
physics which is well-expressed using effective field theory in the intermediate energy regime between the 2HDM and its extension. Taking
this as motivation we analyse Yukawa sector modifications as potential sources to facilitate a strong first-order electroweak phase transition in the early universe. 
While such cancellations reproduce the alignment limit of the 2HDM to maintain consistency with current Higgs data they show up as characteristic deformations of the 2HDM heavy states' phenomenology. 
Not only is this qualitatively different from the scalar sector deformations discussed in Ref.~\cite{Anisha:2022hgv}, but the implied phenomenological
consequences for the LHC are encouraging: Current analysis strategies, whilst remaining robust strategies to lead to discoveries in the future, can overestimate the new physics potential
of exotic Higgs searches in the 2HDM when its deformations to an SFOEWPT are considered as a result of ${O}^{1(12)}_{Qt}$. 

\section*{Acknowledgements}
We thank Matthias Steinhauser for helpful discussions.
This work was funded by a Leverhulme Trust Research Project Grant RPG-2021-031. 
C.E. is supported by the UK Science and Technology Facilities Council (STFC) under grant ST/X000605/1 and the Institute of Particle Physics Phenomenology Associateship Scheme. M.M. is supported by the BMBF-Project 05H21VKCCA.

\appendix
\section{EFT Modifications of Charged Higgs Interactions}
\label{app:coup}
With $\Psi^2\Phi^3$ insertions, the dimension-4 charged Higgs couplings
to the fermions Eq.~(\ref{eq:yuk}) get modified. For the
third-generation fermions, these modifications are 
\begin{multline}
	\mathcal{L}_{\text{charged}}^{\text{dim-6}} = \\ 
	\frac{\sqrt{2}}{v} \left[ \bar{t}\;  \left\{ \left( M_b \,V_{tb}\tan{\beta}  +\frac{v^3 \sin{\beta}}{2 \sqrt{2} \Lambda^2\, }\Big(C_{Qb}^{2(12)} +
	C_{Qb}^{2(21)} +\textcolor{purple}{C_{Qb}^{2(11)}\cot{\beta}+C_{Qb}^{2(22)}\tan{\beta}}\Big)\right) {\text{P}}_{\text{R}}\right.\right.
	\\
	\left. \left.+  V_{tb} \, \left(M_t \, \cot{\beta}   + \frac{v^3 \cos{\beta}}{2 \sqrt{2} \Lambda^2}\left(C_{Qt}^{1(12)} +
	C_{Qt}^{1(21)}+\textcolor{purple} {C_{Qt}^{1(11)} \cot{\beta} + C_{Qt}^{1(22)} \tan{\beta} }\right) \right){\text{P}}_{\text{L}}  \right\} b \; H^+ \right.
	\\
	+ \left. \bar{\nu}_{\tau}  \left(M_\tau \, \tan{\beta}  + \frac{v^3 \sin{\beta}}{2 \sqrt{2} \Lambda^2}\left(C_{L\tau}^{2(12)} +
	C_{L\tau}^{2(21)} +\textcolor{purple}{C_{L\tau}^{2(11)}\cot{\beta}+C_{L\tau}^{2(22)}\tan{\beta}}\right) \right) {\text{P}}_{\text{R}} \, \tau \, H^+ + {\text{h.c.}}\right].
	\\
\end{multline} 
Again for $\Lambda\to \infty$, the standard 2HDM relations of Eq.~\eqref{eq:yuk} are recovered.

\bibliography{draft}

\end{document}